\documentclass[twocolumn,pre,amsmath,amssymb,showpacs]{revtex4-1}
\usepackage{graphicx,color}
\usepackage{amsmath,amssymb,latexsym,amsthm}
\usepackage{mathtools}
\usepackage{mhchem}
\usepackage{tikz}
\usetikzlibrary{chains,shapes,fit,calc,arrows}

\usepackage{color}

\newcommand{\mean}[1]{{\left< #1 \right>}}
\newcommand{\Var}[1]{\text{Var}\, #1 }
\newcommand{\abs}[1]{{\left| #1 \right|}}

\newcommand{\e}{\text{e}}

\definecolor{webgreen}{rgb}{0,.5,0}
\definecolor{webbrown}{rgb}{.6,0,0}
\definecolor{grigio}{rgb}{.85,.85,.85} 
\definecolor{RoyalBlue}{rgb}{0.0, 0.14, 0.4}
\definecolor{skyblue1}{rgb}{0.45,0.62,0.81}
\definecolor{skyblue2}{rgb}{0.2,0.39,0.64}
\definecolor{skyblue3}{rgb}{0.13,0.29,0.53}
\definecolor{scarlet1}{rgb}{0.93,0.16,0.16}
\definecolor{scarlet2}{rgb}{0.8,0,0}
\definecolor{scarlet3}{rgb}{0.64,0,0}

\usepackage{hyperref}
\hypersetup{%
    colorlinks=true, linktocpage=true, pdfstartpage=1, pdfstartview=FitV,%
    breaklinks=true, pdfpagemode=UseNone, pageanchor=true, pdfpagemode=UseOutlines,%
    plainpages=false, bookmarksnumbered, bookmarksopen=true, bookmarksopenlevel=1,%
    hypertexnames=true, pdfhighlight=/O,
    urlcolor=webbrown, linkcolor=RoyalBlue, citecolor=webgreen, 
    pdftitle={Current fluctuations in presence of time-periodic metabolic conditions},%
    pdfauthor={Gianmaria Falasco},%
    pdfsubject={},%
    pdfkeywords={},%
    pdfcreator={pdfLaTeX},%
    pdfproducer={LaTeX REVTeX}%
  }

\begin{document}

\title{Strong current response to slow modulation: a metabolic case-study}
\newcommand\unilux{\affiliation{Complex Systems and Statistical Mechanics, Department of Physics and Materials Science, University of Luxembourg, L-1511 Luxembourg}}
\author{Danilo Forastiere}
\unilux
\author{Gianmaria Falasco}
\unilux
\author{Massimiliano Esposito}
\unilux

\begin{abstract}
  We study the current response to periodic driving of a crucial biochemical reaction network, namely, substrate inhibition. We focus on the conversion rate of substrate into product under time-varying metabolic conditions, modeled by a periodic modulation of the product concentration. We find that the system exhibits a strong nonlinear response to small driving frequencies both for the mean time-averaged current and for the fluctuations. For the first, we obtain  an analytic formula by coarse-graining the original model to a solvable one. The result is nonperturbative in the modulation amplitude and frequency. We then refine the picture by studying the stochastic dynamics of the full system using a large deviations approach, that allows to show the resonant effect at the level of the time-averaged variance and signal-to-noise ratio. Finally, we discuss how this nonequilibrium effect may play a role in metabolic and synthetic networks.
\end{abstract}


\maketitle

\section{Introduction}
Metabolic pathways in living systems generally operate under time dependent conditions.
One the one hand, they experience periodic variations in some environmental input, such as external light, food or energy consumption requirements. On the other, they can display stable endogenous variations in time, such as circadian rhythmicity of hormones allowing for synchronization of biological clocks \cite{rep02}, and calcium oscillations responsible for signal transduction  \cite{ber00, thu12}.
Clearly, these two aspects are closely intertwined, often in a subtle manner. For example, while the role of nutrition as an input for biological clocks is understood to be a crucial aspect \cite{eckel-mahan2013metabolism}, it is still unclear how external stimuli can affect the feedback loop between cytosolic calcium and mitochondrial ATP production \cite{tar12}.

From a modeling perspective, these time dependent behaviors can be seen as driven and autonomous oscillations, respectively. 
In the first case, a modular approach is used that considers a reaction scheme as part of a larger pathway subject to some time-dependent input. The aim is to study how the network topology and nonlinearities may originate a nontrivial chemical output \cite{sam02, sin07}.
In the second, one aims at understanding how long lived oscillations in chemicals' concentration
can stably emerge from the interplay of the intrinsic noise and the (topological and kinetic) features of the chemical reaction network \cite{nov08, cha11,voo19}. 

When reactions are rightfully  described as stochastic Markov processes, a wealth of results and techniques can be employed to tackle the aforementioned questions. Large deviation theory yields the statistics of, e.g., reaction currents in the limit of long observation times or large system size \cite{and08}. Stochastic thermodynamics offers a systematic way to identify the forces driving such currents and the dissipation they entail \cite{rao2018jcp}. By doing so, reaction networks can be examined as chemical machines \cite{penocchio2019thermodynamic}, possibly including the role of information \cite{che19} (e.g. in signal transduction) in the characterization of their performance.
 Finally, interesting phenomena can be expected to accompany driven and autonomous oscillations, such as stochastic resonance \cite{ben81, gammaitoni1998stochastic} and amplification \cite{mck05}, respectively.

Here, we take on such approach of nonequilibrium physics to investigate the effects that time periodic metabolic conditions can have on simple, yet fundamental biochemical motifs. Because of the ubiquity and relative simplicity, we study the metabolic pattern described by the chemical reaction network
\begin{align}\label{eq:sub_inh}
      \ce{E + &S <=>[$k_1$][$k_{-1}$] ES <=>[$k_3$][$k_{-3}$] E + P}\\
  &\ce{ ES + S <=>[$k_2$][$k_{-2}$] ESS}\nonumber
\end{align}
This chemical reaction network goes under the name of substrate inhibition. Because of the presence of the biologically inactive complex ESS, increasing the concentration of substrate S above a certain threshold value results in a decreased yield of the product P \cite{falasco2019negative}. Our analysis focuses on what happens in this network if the concentration of P is subject to a periodic driving, \emph{i.e.} if its concentration is externally fixed depending on chemical or photochemical stimuli that repeat periodically.

\emph{Outline.} We formulate the evolution of the system as a Markov process and discuss the relation with the deterministic rate equations for the kinetics of the chemical network. Assuming that the intermediate species ES is scarcely populated, we obtain an analytic expression for the mean time-averaged current. It displays a resonant behavior at small driving frequencies and it is nonperturbative in the driving amplitude and frequency. To extend the results beyond the average picture, the stochastic dynamics of the full system is studied using large deviations theory. We compare and contrast the quasi-steady-state approximation in the slow driving limit with the general approach for large deviations of periodically driven systems \cite{verley2013modulated, barato2018current}. We conclude exploring the possible implications of this resonance for metabolic and synthetic chemical reaction networks.

\section{Substrate inhibition}
\label{sec:substrate}
We recall some important facts of the analysis of substrate inhibition at the steady-state, i.e. in absence of a periodic modulation of the substrate, as thoroughly discussed in  \cite{falasco2019negative}.

\subsection{Stochastic dynamics}
We define the number of molecules of the dynamical species E, ES, ESS as $n_{\text{E}}, n_{\text{ES}}$ and  $ n_{\text{ESS}}$, respectively. We denote by $s$ and $p$ the concentrations of the chemostatted, i.e. externally controlled, species (interpreted, respectively, as the substrate S and the product P of the enzymatic reaction). The stochastic dynamics of the species E, ES, ESS is given by 
\begin{align}\label{process}
d n_i = \sum_{\rho=1}^{3} (\nu_{-\rho i}dN_{-\rho}-\nu_{+\rho i}dN_{+ \rho})
\end{align}
where $dN_{\rho}$ are the random numbers of reactions $\rho$ in the infinitesimal time $dt$, i.e. independent Poisson variables with intensity $W_\rho(n)=\sum_i \nu_{\rho i} k_\rho n_i$. The stoichiometric coefficient $\nu_{\rho i}= \{\pm 1, 0\}$ is the number of reactants of species $i$ involved in the reaction $\rho$. A description equivalent to \eqref{process} is given by the chemical master equation \cite{gil92}.
 
However, rather than specifying how many molecules of each species are present, we can specify in which chemical state $i$ (E, ES, or ESS) a single molecule is. In other words, since the network is linear, the dynamics of $M$ molecules is equivalent to $M$ independent replicas of the dynamics of a single molecule. Or, more formally, the stochastic state $i(t)$ has the same statistics of $n_i(t)/M$. This simplifies considerably the description, since the dynamics can be now described in the 3-dimensional state of chemicals---instead of the much larger space of occupation number---by the master equation for the probability of the state $i$ \cite{laz19}. By construction, the latter coincides with the rate equations, that is the ensemble average of \eqref{process},
\begin{equation}
\begin{aligned}\label{rateeq}
    \dot{c}_{\text{E}} &= (k_{-1}+k_3) c_{\text{ES}}(t) - (k_1 s+k_{-3} p) c_{\text{E}}(t)  \\
       \dot{c}_{\text{ES}} &= (k_1 s+k_{-3} p) c_{\text{E}}(t) - K c_{\text{ES}}(t) + k_{-2} c_{\text{ESS}}(t) \\
    \dot{c}_{\text{ESS}}&= k_2 s\, c_{\text{ES}}(t)-k_{-2} c_{\text{ESS}}(t)
  \end{aligned}
\end{equation}
with $c_i:=\mean{n_i}/V$ the macroscopic concentrations in some reference volume $V$, and
\begin{align}
K:=k_{-1}+k_2s+k_3\,.
\end{align}

\subsection{Steady-stated  analysis}
The unique steady-state distribution for this model, given  $M/V=c_{\text{E}}(0)+c_{\text{ES}}(0)+c_{\text{ESS}}(0)$ as initial total concentration, is 
\begin{align}
  \begin{aligned}\label{eq:ss}
    c_{ \text{E}}^{\text{ss}}&= \frac{M }{V d}k_{-2}(k_{-1}+k_3)\,,  \\
    c_{\text{ES}}^{\text{ss}}&= \frac{M }{V d} k_{-2}(k_{-3}p+k_1s)\,,  \\
    c_{ \text{ESS}}^{\text{ss}}&=\frac{M }{V d}k_2s(k_{-3}p+k_1s) ,
    \end{aligned}
  \end{align}
  where
  \begin{align}
d:=k_{-2}(k_{-1}+ k_3) +(k_{-2}+ k_2s) (k_1s+k_{-3}p )\,.
  \end{align}

In the following, we will need to compute the cumulants of the current 
\begin{align}
J(t) = \int_0^t  (dN_{+3}(t)- d N_{-3}(t)) \label{eq:current}
\end{align}
that is the difference in the counting processes relative to the reaction $\rho=+3$ and $\rho=-3$. For long times, this is the current that transforms the substrate into product, as it is shown in details in Appendix \ref{app:pathint}. Its cumulants are found introducing a large deviation formalism that relies on the use of Gardner-Ellis theorem \cite{tou09}. We define the scale-cumulant generating function of $J$ as
\begin{align}
  \text{scgf}(\lambda)=\lim_{t\to \infty}\frac{1}{t} \ln \mean{\e^{\lambda \cdot J(t)}}.\label{eq:scgf}
\end{align}
For the Markov property of the stochastic model associated to the rate equations \eqref{rateeq}, Eq.~\eqref{eq:scgf} can be obtained as the dominant eigenvalue of the \emph{tilted generator}
\begin{align}
\mathcal{T(\lambda)}=
  \begin{pmatrix}
    -(k_1s+k_{-3}p) & k_{-1}+k_3\e^{\lambda}& 0 \\
    \e^{-\lambda}k_{-3}p + k_1s& -K & k_{-2}\\
    0 & k_2s & -k_{-2}\\ 
  \end{pmatrix}\,.
  \label{eq:tiltedgen}
\end{align}
As the name suggests, \eqref{eq:tiltedgen} is a modified generator of the state dynamics, which `counts' $\pm \lambda$ every time the reaction $\rho=\pm 3$ takes place. It reduces to the dynamics \eqref{rateeq} for $\lambda =0$.
The eigenvalue equation for this model is an algebraic equation of degree 3 in $\lambda$, that can be solved analytically.
Using the scaled-cumulants generating function it is possible to obtain all the cumulants of the current $J$ at steady-state. In particular, the mean steady-state current is 
\begin{align}
    J_{\text{ss}}&=\left.\frac{\partial \,\text{scgf}}{\partial \lambda}(\lambda)\right|_0= \frac{M}{V d} k_{-2} (k_1 k_3 s-k_{-1} k_{-3} p)\,.
\end{align}

\begin{figure}
  \includegraphics[width=\columnwidth]{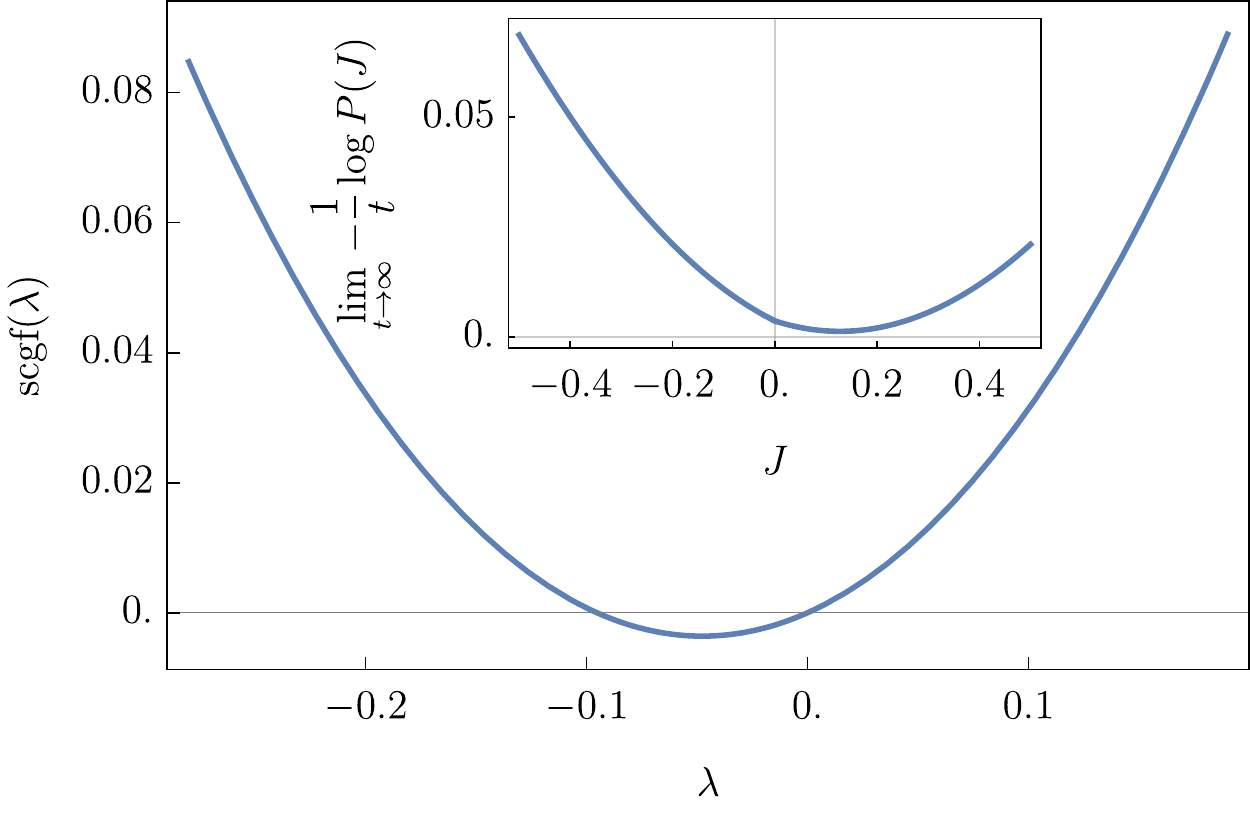}
  \caption{Scaled-cumulant generating function for the integrated current $J$ when the system is at the steady-state defined by Eqs. \eqref{eq:ss}. \emph{Inset:} The corresponding  scaled logarithm of the probability $P(J)$ of the integrated current $J$. Parameters used are (arbitrary units) $k_1=30$, $k_{-1} = 300$, $k_2= 200$, $k_{-2} =2$, $k_3 = 200$, $k_{-3} =10$, $p_0= 1$, $s=0.55$, $M/V = 1$.}
  \label{fig:scgf}
\end{figure}
An important symmetry of the dynamics is  the \emph{steady-state fluctuation theorem} \cite{lebowitz1999gallavotti}, as
\begin{align}
\text{scfg}(\lambda)=\text{scfg}(\mathcal{A}-\lambda) \label{eq:fluctuation_thm}
\end{align}
Here $\mathcal{A}$ is the (dimensionless) \emph{chemical affinity} of the chemical network \cite{rao2018jcp}
\begin{align}
  \mathcal{A}:= \ln\frac{sk_1k_3}{p_0 k_{-1}k_{-3}}\,, \label{eq:affinity}
\end{align}
$p_0$ being the reference value of the product concentration around which we will consider periodic modulation.
The steady-state entropy production of this system can be written in terms of the steady-state current and affinity, as
\begin{align}
  \label{eq:ss_epr}
\dot\sigma = J_{\text{ss}}\, \mathcal{A}\,. 
\end{align}
Since $\dot\sigma$ is positive definite, the sign of $\mathcal{A}$ determines the sign of the steady-state current.

\subsection{Periodic driving}
When one or more of the chemostatted species vary periodically in time due to an external driving of period $T$, the natural quantities to study  are the time-averaged cumulants of the current \eqref{eq:current}, e.g.

\begin{align}
  \mean{\overline{J}} &:= \lim_{n \to \infty} \frac{1}{n T} \int_0^{nT} d t\, \mean{J(t)} \,,\\
  \overline{\text{Var} J} &:= \lim_{n \to \infty} \frac{1}{n T} \int_0^{nT} d t\, \text{Var}J(t)\,.
\end{align}
These cumulants can still be obtained from the scaled-cumulant generating function \eqref{eq:scgf}, but the property that the latter is the dominant eigenvalue of the tilted-generator is no more valid. The correct procedure to compute it is then to use the formalism developed in \cite{barato2018current, verley2013modulated} and exploited in Section \ref{sec:floquet}. Before dealing with this general approach,  we will  obtain the mean values of the dynamical observables in the time-dependent case via a direct method. We point out that in the general case of driving with an arbitrary protocol, the symmetry \eqref{eq:fluctuation_thm} does not hold anymore and should be substituted by the generalization in \cite{rao2018jcp}, that includes periodic driving, as well as boundary contributions.

The periodic driving we will consider is the time variation of the chemostatted concentration $p$ according to the protocol (with $0< \gamma <1$)
\begin{align}
    p(t)&=p_0\left(1+\gamma \sin (\Omega t) \right) \label{eq:protocol}\,.
\end{align}
The original ODE system becomes non-autonomous and reads
\begin{align}
 & \dot{c}_{\text{E}}   = (k_{-1}+k_3) c_{\text{ES}}(t) - (k_1 s+k_{-3} p(t)) c_{\text{E}}(t) \,, \label{eq:ode1}\\
 & \dot{c}_{\text{ES}}  = (k_1 s+k_{-3} p(t)) c_{\text{E}}(t) - K c_{\text{ES}}(t) + k_{-2} c_{\text{ESS}}(t)\,, \label{eq:ode2}\\
 & \dot{c}_{\text{ESS}} = k_2 s\, c_{\text{ES}}(t)-k_{-2} c_{\text{ESS}}(t)\,. \label{eq:ode3}
\end{align}
Floquet theory implies that the trajectories relax to a periodic steady-state, independent on the initial condition. However, it is not possible to solve this system given an initial condition for any choice of the rate constants. In the following we will give conditions under which it is possible to map this 3-state model to a low dimensional one that is solvable for any value of the parameter $\gamma$.

The thermodynamics of this model can be formulated in terms of the entropy production \cite{rao2018jcp} in the periodic steady-state, that averaged on a period reads
\begin{align}
  \overline{ \dot \sigma}= &\frac{1}{T} \int_0^T dt  \mean{J(t)} \mathcal{A}(t)  - \frac{1}{T} \int_0^T dt\,  \dot{p}(t) \mathcal{A}(t) \,.\label{eq:epr}
\end{align}
Compared to Equation \eqref{eq:ss_epr}, in presence of time-dependent driving an additional  contribution must be considered other than the product of current and  affinity. This modification in the structure of the entropy production means that the sign of the time-averaged current $\mean{\overline{J}}$ is no more constrained by the Second Law to be equal to the sign of the affinity $\mathcal{A}$ of the corresponding steady-state. 
We will use the notation $\mathcal{A}(t)$ for the instantaneous chemical affinity
\begin{align}
  \mathcal{A}(t):= \ln\frac{sk_1k_3}{p(t) k_{-1}k_{-3}}\,,
\end{align}
while the value correponding to the average product concentration $p_0$ will be denoted by $\mathcal{A}=\mathcal{A}(0)$.

\section{Analytic expression for $\mean{\overline{J}}$}
\label{sec:reduction}
In biological systems, the concentration of the intermediate complex ES decays rapidly into one of the other species because of the values of the kinetic rates. We can exploit this feature to obtain a solution of the system \eqref{eq:ode1}-\eqref{eq:ode3} by coarse-graining the intermediate complex, leading to an analytic expression for $\mean{\overline{J}}$.
This coarse-graining is possible when the condition $K\gg1$ is verified, and under this hypothesis the system exhibits a time-scale separation that allows us to project the three states model onto an equivalent two states one.
Inspired by the Mori-Zwanzig approach \cite{zwanzig1973nonlinear, rubin2014memory}, we write the formal solution of Eq. \eqref{eq:ode2} as 
\begin{align}
  c_{\text{ES} }(t) &= \int_0^t dt' \e^{-K (t-t')}\{ (k_1 s + k_{-3} p(t')) c_{\text{E}}(t') \\
  & \qquad \qquad \qquad \qquad + k_{-2} c_{\text{ESS}}(t') \} \nonumber \\ 
                    &\simeq \frac{1}{K}\left\{\left(k_1s + k_{-3} p(t)\right) c_{\text{E}}(t) + k_{-2} c_{\text{ESS}}(t)\right\}
          \label{eq:es_solution}            
\end{align}
where we used the approximation $\e^{-K(t-t')}\simeq 0$ for all $t'\neq t$ to neglect all the contribution of the integrand but at time $t'=t$. Setting the initial condition $c_{\text{ES} }(0)=0$ we have considered the concentration of ES after a sufficiently long  relaxation time. From the requirement that the integrand is approximately constant on the interval where the exponential is significantly nonzero, we obtain a condition for the time-scale separation, expressed in terms of the minimum frequency $\frac{\Omega}{2}$  of the periodic solutions for  $c_{\text{E}}$ and $c_{\text{ESS}}$ (obtained using Floquet theory) as $\frac{\Omega}{2} \ll K$.

This projection is useful when combined with the conservation law for the total concentration of molecules $M/V=c_{\text{E}}(t)+c_{\text{ES}}(t)+c_{\text{ESS}}(t)$ at any time $t$. In fact, plugging this conservation law into Eq. \eqref{eq:es_solution} we are able to write down a single first-order differential equation (with time dependent coefficients due to the time dependence in the chemostat $p$) that describes the full dynamics of the chemical pathway in the appropriate regime. 
Introducing the quantities
\begin{align}
    C&=\frac{M}{V}\frac{k_{-2} (k_{-1}+k_3)}{K+k_{-2}}\,,\\
    R&=\frac{d}{K+k_{-2}}\,,\\
    f(t)&=-\gamma  \frac{  k_{-3} p_0 (k_2 s+k_{-2})}{K+k_{-2}}  \sin ( \Omega t )\,, \label{eq:oscillating_term}
\end{align}
the final ODE reads
\begin{align}
  \dot{c}_{\text{E}}&=C - (R - f(t)) c_{\text{E}}(t)\,. \label{eq:1dmodel}
\end{align}
This differential equation is still not solvable in general, but we can proceed as above to obtain conditions on the time-scales such that we can write down an explicit solution. Neglecting the initial condition, we can integrate both sides of Eq. \eqref{eq:1dmodel} and, under the condition $ \abs{\int_\tau^tf(s) d s} \ll 1$, take the dominant contribution to the integral  to obtain
\begin{align}
    c_{\text{E}}(t) &= C \int_0^t d \tau \e^{-R(t-\tau)} \exp\left\{\int_\tau^t f(s)  d s \right\} \label{eq:2int}\\
           &\simeq C \int_0^t d \tau \e^{-R(t-\tau)} \left(1+\int_\tau^tf(s) d s \right)\,. \label{eq:2int_approx}
\end{align}

We can provide sufficient conditions under which the requirement $\abs{\int_\tau^tf(s) d s} \ll 1$ is fulfilled. Clearly, the expansion \eqref{eq:2int_approx} holds  if \eqref{eq:oscillating_term} is  much smaller than one uniformly, \emph{i.e.} if
\begin{align}
\gamma  \frac{  k_{-3} p_0 (k_2 s+k_{-2})}{K+k_{-2}}\ll 1\,.
\end{align}
On the other hand, even if the latter term is of order unity, equation \eqref{eq:2int_approx} remains valid if the driving frequency  is much smaller than the rate shaping the interval on which the kernel $\e^{-R(t-\tau)}$ is significantly different from zero, that is if
\begin{align}
\Omega \ll R\,.
\end{align}
Under this condition the term \eqref{eq:oscillating_term} would be approximatively zero for all the relevant integration times.

The solution for $c_{\text{E}}$ in the periodic steady-state, valid for any value of $\gamma \in (0,1)$, given the previous conditions on the rates, finally is
\begin{align}
  c_{\text{E}}^{\text{ps}}(t)&=c_{\text{E}}^{\text{ss}}\left(1+\gamma \delta c_{\text{E}}^{\text{ps}}(t)\right)
\end{align}
with the definitions
\begin{align}
  \delta c_{\text{E}}^{\text{ps}}(t)&=  -  a \frac{d \sin ( \Omega t ) -\Omega  (K+k_{-2}) \cos ( \Omega t)}{\Omega ^2 (K+k_{-2})^2+d^2} \,,\\
  a&:= k_{-3} p_0 (k_2 s+k_{-2}) \,.
\end{align}
The correction to the deterministic current averaged on a period,
\begin{align}
\overline{\Delta J} :=\mean{\overline{J}}- J_{\text{ss}}  \,,
\end{align}
can be then expressed as
\begin{align}
  \overline{\Delta J} &= \gamma^2b \frac{M}{V}\frac{ k_{-2} k_{-3}^2 p_0^2 (k_{-1}+k_3) (k_2 s+k_{-2})}{
                        2 \left( d^2+  K^2 \Omega ^2\right)}\,, \label{eq:current_correction}\\
  b &:=\left( 1-\frac{k_3}{K+k_{-2}}\right) >0\,.
\end{align}

\begin{figure}
  \includegraphics[width=\columnwidth]{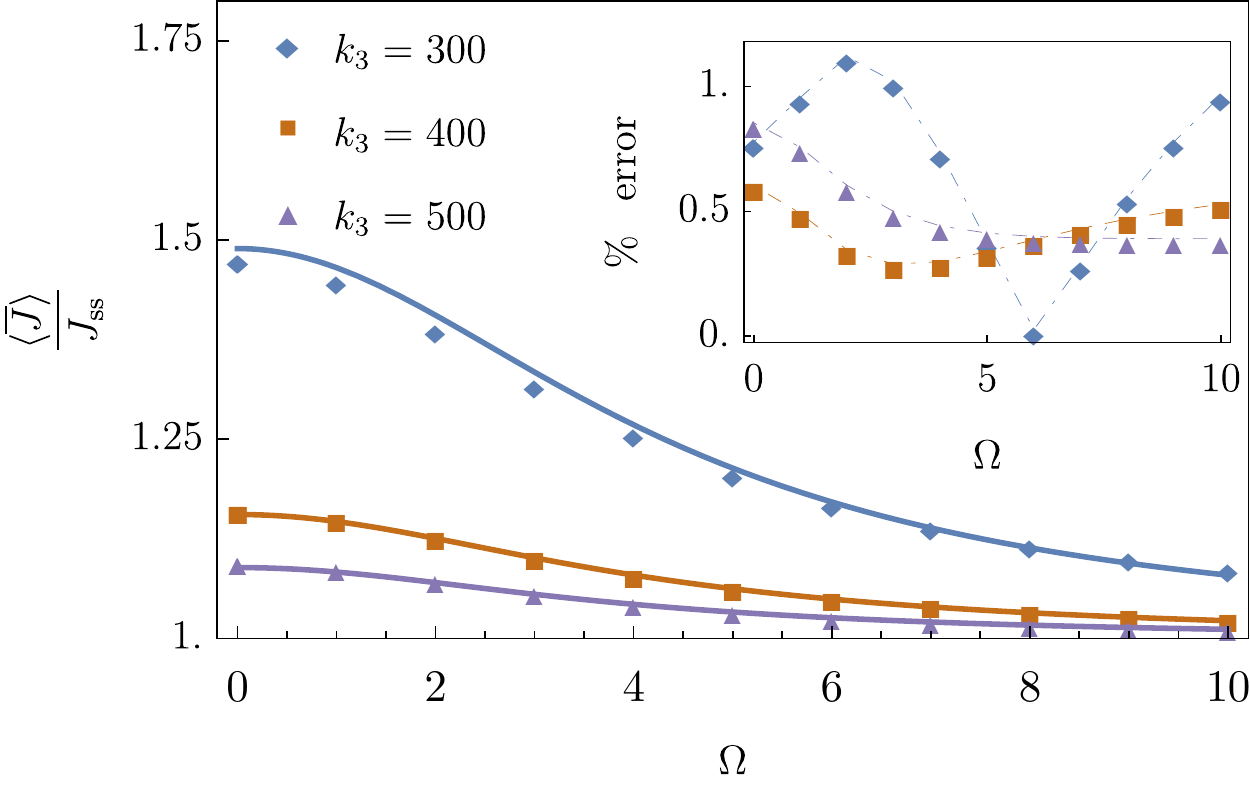}
  \caption{Time-averaged current  $\mean{\overline{J}}$ normalized by its steady-state value $J_{\text{ss}}$ plotted as a function of the driving frequency $\Omega$, for the case $\mathcal{A} > 0$. The solid line gives the theoretical prediction while the diamonds represent the results of the numerical integration of the Eqs. \eqref{eq:ode1}-\eqref{eq:ode3}. \emph{Inset:} relative error between the theory and the numerical results. The graph is obtained for $s=0.4$, $\gamma=0.8$. The other parameters are the same of Figure \ref{fig:scgf}, if not stated differently.}
  \label{fig:positiveDeltaJ} 
\end{figure}

\begin{figure}
  \includegraphics[width=\columnwidth]{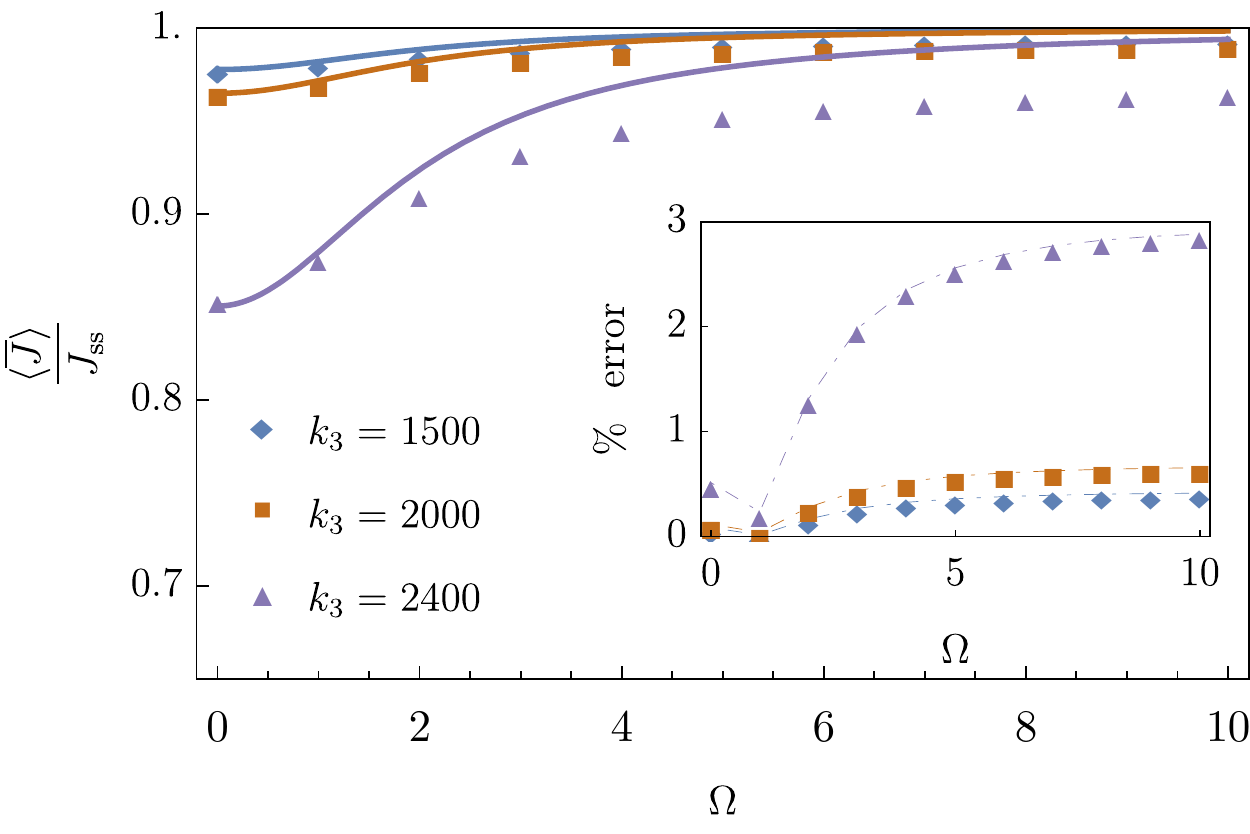}
  \caption{Time-averaged current $\mean{\overline{J}}$ normalized by its steady-state value $J_{\text{ss}}$ plotted as a function of the driving frequency $\Omega$, for the case $\mathcal{A} < 0$. The solid line gives the theoretical prediction  while the diamonds represent the results of the numerical integration of the Eqs. \eqref{eq:ode1}-\eqref{eq:ode3}. \emph{Inset:} relative error between the theory and the numerical results. The graph is obtained for $s=0.04$, $\gamma=0.8$. The other parameters are the same of Figure \ref{fig:scgf}, if not stated differently. }
  \label{fig:negativeDeltaJ}
\end{figure}

While the contribution $\overline{\Delta J}$ of the time periodic protocol is always positive, the sign of the steady-state current is governed by the sign of the chemical affinity $\mathcal{A}$.
When the affinity is positive, the chemical system is converting the substrate S into product P, and i.e. it is possible to identify parameters for which the increase in amplitude of the mean time-averaged current as a function of the frequency is significant, shown in Fig. \ref{fig:positiveDeltaJ}. Note that the maximum of $\overline{\Delta J}$ is obtained at $\Omega =0$. Since for this protocol $\Omega=0$ means that no perturbation is applied, the appearance of the maximum should be interpreted as a discontinuity in the time average of $J$ reflecting the fact that the latter is a highly nonlinear functional of the driving protocol. Intuitively, this can be understood as an effect of the growth of the integration interval of the time average as the frequency goes to zero. This heuristic explanation is confirmed by the numerical results at very low frequencies shown in Figures \ref{fig:positiveDeltaJ} and \ref{fig:negativeDeltaJ}.

When $\mathcal{A}<0$ is negative in the substrate-inhibition scheme, we can consider a dynamically equivalent product-inhibition reaction
 \begin{align}\label{eq:prod_inhib}
  \ce{E + &S <=>[$k_{-3}$][$k_3$]ES <=>[$k_{-1}$][$k_1$] E + P }  \\
  &\ce{ ES + P <=>[$k_2$][$k_{-2}$]ESP}\nonumber
 \end{align}
 with affinity $\mathcal{A}'=-\mathcal{A}>0$.
In this case, the interpretation of the terms in Eqs. \eqref{eq:ode1}-\eqref{eq:ode3} changes, and $p(t)$ (resp., $s$) is now the concentration of the substrate S (product P). As displayed in Figure \ref{fig:negativeDeltaJ}, it is possible to choose the parameters for which the behavior of the average current is significantly less than the current at steady-state.

We conclude this section providing a simple interpretation of this phenomenon. Assume that the dynamics is such that increasing P, in the first half-period, leads to an increase of ES and ESS without delay. As a result of the conservation law $c_{\text{E}}+ c_{\text{ES}}+c_{\text{ESS}}= M/V$, the concentration of E decreases during this first part of the protocol. Analogously, in the second half-period the concentrations of P, ES, ESS are decreasing simultaneously, while E is increasing. By this argument, we see that there is a phase shift of $\pi$ between $p(t)$ and $c_{\text{E}}(t)$. This picture clearly breaks down when the driving frequency is too high and the response is not instantaneous anymore. In this limit, \emph{i.e.} when $\Omega \to \infty$, the system's response is slow compared to the driving, and the concentration of P can be taken as constantly  equal to its average value $p_0$ (for a numerical proof of this fact, see \emph{e.g.} the high frequency region of the inset of Figure \ref{fig:ld_variance} in Section \ref{sec:floquet}). 
Under the hypothesis of moderate driving frequency, thus, there is no delay in the (nonlinear) response of the instantaneous current to the time dependent perturbation, that takes the form
\begin{align}
  J(t) &= k_3 s \,c_{\text{ES}}(t) - k_{-3} p(t) c_{\text{E}}(t) \label{eq:time_dep_current}\\
       &= J_{\text{ss}} + \gamma J_1 \sin(\Omega t)  + \gamma^2 J_2 \sin^2(\Omega t) \label{eq:perturbative_expansion}
\end{align}
with $J_2 > 0$. After averaging over a period, the order $O(\gamma)$ term cancels out, so that the response is of order $O(\gamma^2)$, i.e. fully nonlinear.
Under the condition $\Omega \ll K$, and using the conservation law, the concentration $c_{\text{ES}}$ can be expressed in terms of  $c_{\text{E}}$. The second-order contributions in \eqref{eq:time_dep_current} are both proportional, with a minus sign, to $ p(t) c_{\text{E}}(t)$, so that the sign of $J_2$ is positive because of the phase shift existing between the concentrations of P and E. 

\begin{figure}
  \includegraphics[width=0.87\columnwidth]{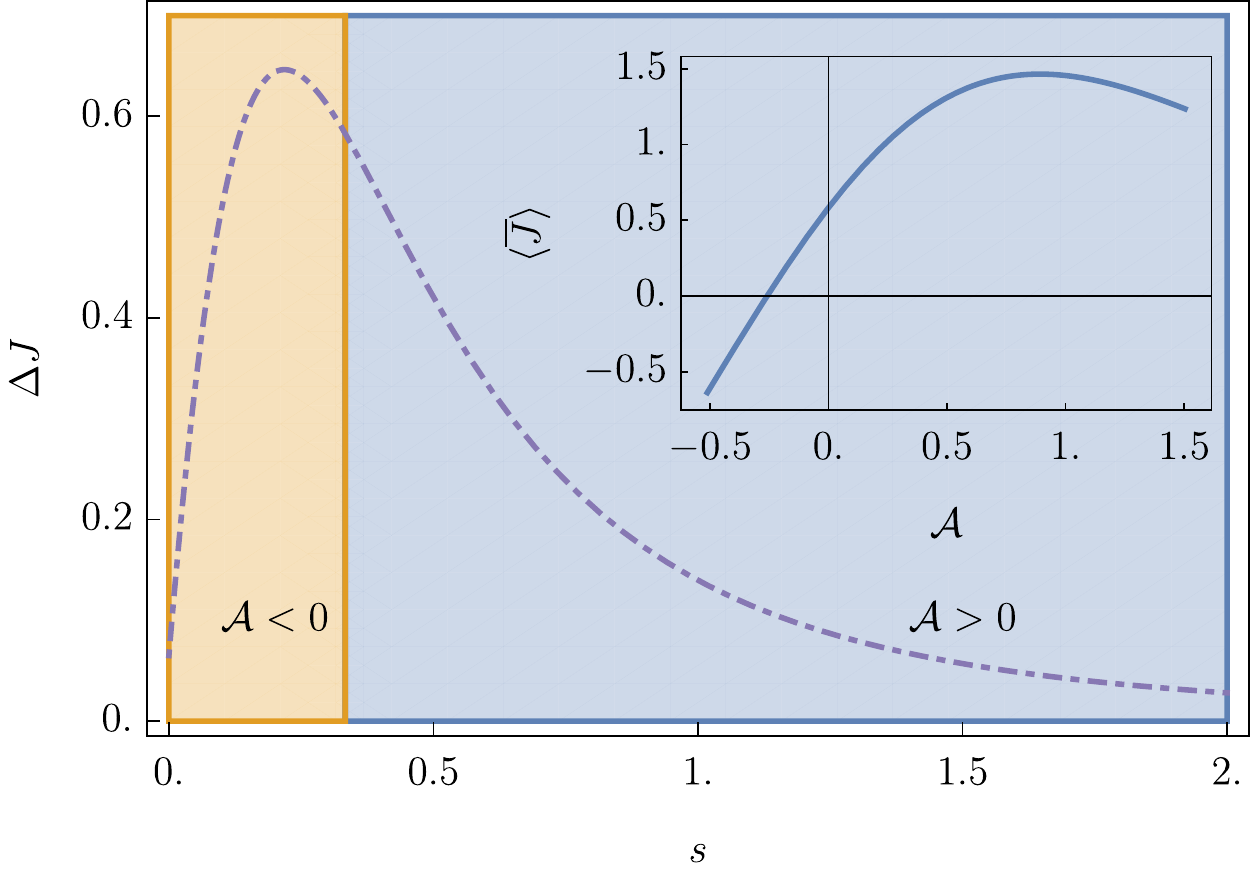}
  \caption{Difference between the average and the steady-state current $\Delta J$ as a function of the substrate concentration $s$. This plot is obtained for $\gamma=0.8$ and $\Omega=1$. The other parameters are the same of Figure \ref{fig:scgf}. \emph{Inset:} Total average current $\mean{\overline{J}}$ as a function of the steady-state affinity.}\label{fig:opt}
\end{figure}

Formula \eqref{eq:current_correction} is a chemical analog to known results in the literature on stochastic resonance \cite{ben81, gammaitoni1998stochastic}. 
Because the system is an open chemical network, the value of the effective reaction rates depend on the (average) concentrations of P and S. For any given driving frequency, is then possible to maximize the increase (or decrease) of the time-averaged current by changing the chemostats' concentrations accordingly. Critical values of $s$ are found by solving (for given values of the kinetic rates and of $p_0$)
\begin{align}\label{eq:optimization}
  \frac{\partial \overline{\Delta J}}{\partial s} = 0
\end{align}
subject to the constraint that the reference affinity $\mathcal{A}(s)$ is either positive or negative, if we want the current to have a definite sign (see Figure \ref{fig:opt}). We point out that our treatment differs from the classic picture in at least two main ways. First, the typical quantities considered in models of stochastic resonance are state observables, while in our case we are interested in a current. Second, since such current is time-averaged over a period, the linear response is identically zero after averaging, so that the correction to the steady-state current is a genuinely nonlinear effect.

The inset of Figure \ref{fig:opt} shows that the system can also exhibit negative absolute response to the steady-state affinity \cite{machura2007absolute} --- \emph{i.e.} the total current $\mean{\overline{J}}$ can be positive even for negative values of $\mathcal{A}$. This effect does not violate any thermodynamic contraint, like the second law, because the entropy production \eqref{eq:epr} is formulated in terms of the instantaneous current and affinity, rather than the time-averaged current and steady-state affinity.  

\section{Current fluctuations with periodic driving}
\label{sec:fluctuations}
We now want to address the question of how the time-periodic driving modifies the current fluctuations of this exemplary enzymatic reaction, and how these changes combine with the previous phenomenology of the time-averaged current to alter the transduction properties of the biochemical system. To enable comparison  with the results of Section \ref{sec:reduction}, we will continue to consider the condition $K\gg 1$ valid for the kinetic rates. However, we remark that the techniques of this Section do not rely on this hypothesis.
\subsection{Quasi-steady-state approximation (QSS)}
\label{sec:qss}

\begin{figure}
  \includegraphics[width=\columnwidth]{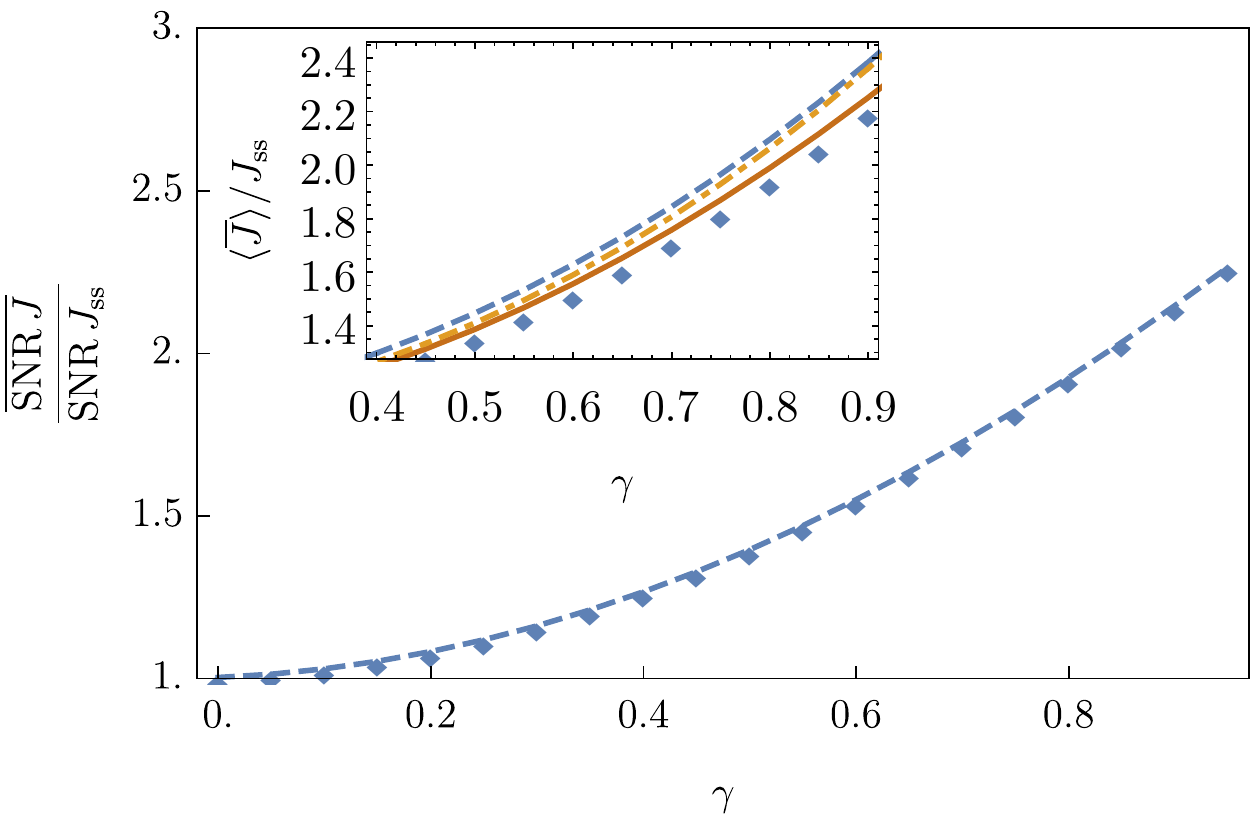}
  \caption{ Time-averaged signal-to-noise ratio with driving $\overline{\text{SNR}\,J}$ normalized by the corresponding steady-state values $ \text{SNR}\,J_{SS}$ as a function of the driving amplitude $\gamma$. Diamonds are obtained by numeric integration of the integral \eqref{eq:qss_fluctuations}. The dashed line is a guide to the eye. The time-averaged variance $\overline{\Var J}$ (not shown) remains approximatively equal to the corresponding steady-state value $ \Var{ J_{SS}}$. The parameters are as in Figure \ref{fig:scgf}. Inset: Time-averaged current normalized by its steady-state value. Diamonds mark points obtained from the numerical integration of the rate equations with $\Omega=1$, to be compared with the prediction of Eq. \eqref{eq:current_correction} (solid line), the QSS approximation in Eq. \eqref{eq:qss_fluctuations} (dashed) and the QSS theory corrected with the inclusion of the geometric correction \cite{sin07}, as discussed in Appendix \ref{app:pathint} (dot-dashed). Other parameters as in Figure \ref{fig:scgf}.  }
  \label{fig:qss}
\end{figure}
The quasi-steady-state approximation postulates that, if the driving protocol is infinitely slow, for any given value of the driving protocol the cumulants are given by their corresponding instantaneous steady-state value. These values are in turn computed using the scaled cumulant generating function obtained as the dominant eigenvalue of the \emph{tilted generator} in Eq. \eqref{eq:tiltedgen}. The first two cumulants of $J$  are thus expressed as the average over one period of their instantaneous values, that is
\begin{align}
  \begin{aligned}
  \mean{\overline{J}} &\simeq \frac{1}{T} \int_0^T d t   \left.\frac{\partial\, \text{scgf}}{\partial \lambda}(\lambda)\right|_{\lambda=0, p(t)} \\
  \overline{\Var J} &\simeq \frac{1}{T} \int_0^T d t   \left.\frac{\partial^2 \text{scgf}}{\partial \lambda^2}(\lambda)\right|_{\lambda=0, p(t)} \,.
  \label{eq:qss_fluctuations}
  \end{aligned}
\end{align}
The condition for this to be valid is that all the rates should be much bigger than the periodicity, that is $\min_i{\{k_{\pm i}\}}\gg \Omega $. Clearly, in this regime, the fluctuations are independent on the frequency. In fact, for this particular protocol
\begin{align}
  \left.\dot p(t)\right|_{p(t)=p}= \pm \Omega p_0 \gamma \sqrt{1 - \left(\frac{p-p_0}{p_0\gamma}\right)} =: p'_{\pm} (p)
\end{align}
so that the integral defining the average variance in equation \eqref{eq:qss_fluctuations} can be written as
\begin{align}
   &\frac{\Omega}{2 \pi} \int_0^{2 \pi/\Omega} d t   \frac{\partial^2 \text{scgf}}{\partial \lambda^2}(\lambda)\bigg|_{\lambda=0, p(t)} \\
  &=\frac{\Omega}{2 \pi} \bigg( \int_{p_{\text{min}}}^{p_{\text{max}}} d p \, \frac{1}{p'_{+}(p)} \frac{\partial^2 \text{scgf}}{\partial \lambda^2}(\lambda)\bigg|_{\lambda=0, p} \nonumber\\
  &\qquad  + \int_{p_{\text{max}}}^{p_{\text{min}}} d p \, \frac{1}{p'_{-}(p)} \frac{\partial^2 \text{scgf}}{\partial \lambda^2}(\lambda)\bigg|_{\lambda=0, p}\bigg)\,.\label{eq:qss_path}
\end{align}
The factor $\Omega$  coming from the time average is canceled out by the one contained in the expression of the derivative, resulting in a frequency-independent value of the average variance.  By the same argument, the net effect on the average current due to the driving frequency is expected to be constant in $\Omega$.

We use the time-averaged signal-to-noise ratio defined  by 
\begin{align}\label{snr}
\overline{\text{SNR}\, J }:= \frac{\mean{\overline{J}}}{\left(\overline{\Var{J}}\right)^{\frac{1}{2}}}
\end{align}
to compare the \emph{precision} of the system with driving to its functioning at steady-state. In Figure \ref{fig:qss} we show \eqref{snr} calculated by means of \eqref{eq:qss_fluctuations} as a function of the driving amplitude $\gamma$. We note that, even though the protocol by which the chemostat changes in time is time-symmetric, the net effect on the current is non-vanishing as the chemical affinity \eqref{eq:affinity} driving the current changes in an asymmetric way.

We showed here that even if the quasi-steady state (possibly supplemented by the geometric correction discussed in Appendix \ref{app:pathint}) can reproduce the qualitative behavior of the current's cumulants for vanishing frequency, it is not enough to account for the frequency effects already at the level of the time-averaged current (see the Inset of Figure \ref{fig:qss}) for small but finite frequency. 

\subsection{Arbitrary driving frequency}
\label{sec:floquet}
To surmount the intrinsic limits  of the quasi-steady-state approach  we implemented the general formalism of large deviations for periodically driven systems described in \cite{barato2018current} for the case of our model system.  
The correct scaled-cumulant generating function for the long-time periodic steady-state is given by
\begin{align}
\text{scgf}(\lambda, \Omega) := \lim_{m\to \infty}\frac{1}{m} \ln \mean{\e^{\lambda \cdot J\left( \frac{2 \pi}{\Omega} m \right)}} = \frac{2 \pi}{\Omega} \ln \mu(\lambda, \Omega) \label{eq:periodicscgf}\,,
\end{align}
in which $\mu(\lambda, \Omega)$ is the maximum eigenvalue  of the monodromy matrix $\mathcal{M}(\lambda)=\overleftarrow{\exp}\left\{ \int_0^T d \tau \, \mathcal T(\tau,\lambda)\right\}$, and the time-dependent generator $\mathcal{T}(t,\lambda)$ is obtained from \eqref{eq:tiltedgen} by allowing the product concentration to vary in time according to a protocol $p(t)$ characterized by a frequency $\Omega$. As in \cite{barato2018current}, we have represented the monodromy matrix as a time-reversed ordered exponential.
To make the computation easier, we substitute the protocol \eqref{eq:protocol} with a piece-wise constant protocol switching  between $p_{\text{min}}=p_0(1-\frac{1}{2}\gamma)$ and $p_{\text{max}}=p_0(1+\frac{1}{2}\gamma)$ every half-period $T/2$ \cite{verley2013modulated}.

\begin{figure}
  \includegraphics[width=\columnwidth]{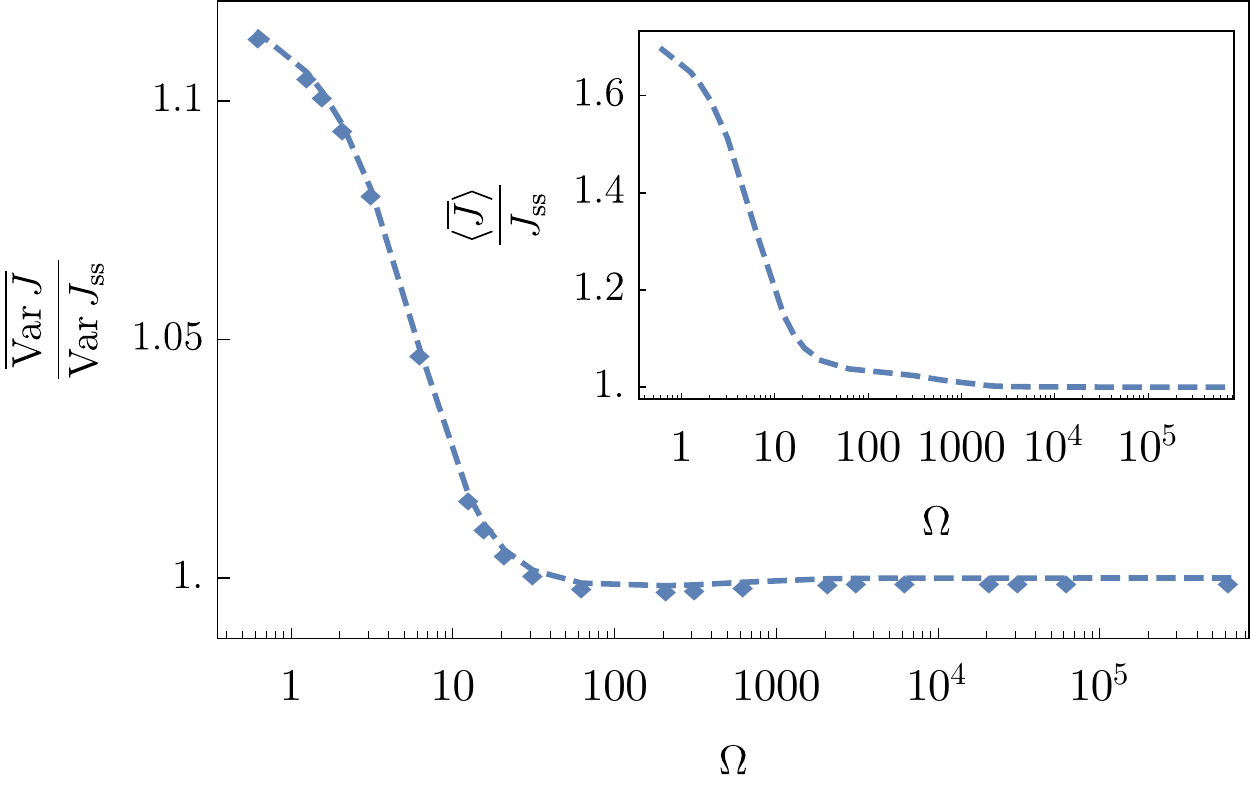}
  \caption{Time-averaged variance with driving $\overline{\Var J}$ normalized to the corresponding steady-state value $ \Var{ J_{SS}}$ as a function of the driving frequency $\Omega$. Diamonds mark points obtained numerically, the dashed line is a guide to the eye. \emph{Inset:} Normalized time-averaged current $\overline{J}/ J_{SS}$. The current resulting from the driving with the step-wise constant driving agrees qualitatively with the one computed with the method of Section \ref{sec:reduction}. Parameters for this plot are the same of Figure \ref{fig:scgf}, and $\gamma=1$. }
  \label{fig:ld_current}
  \label{fig:ld_variance}
\end{figure}

To assess the stability of the results with respect to the choice of the simplified protocol, we studied the time-averaged current $\overline{J}$ for the same choice of kinetic parameters and initial concentrations used for Figure \ref{fig:positiveDeltaJ}. The inset in Figure \ref{fig:ld_current} shows that the qualitative behavior of the current is the same as the one obtained using the protocol \eqref{eq:protocol}, with a significant peak around $\Omega=0$, while the effect disappears at very high frequencies.
\begin{figure}
   \includegraphics[width=\columnwidth]{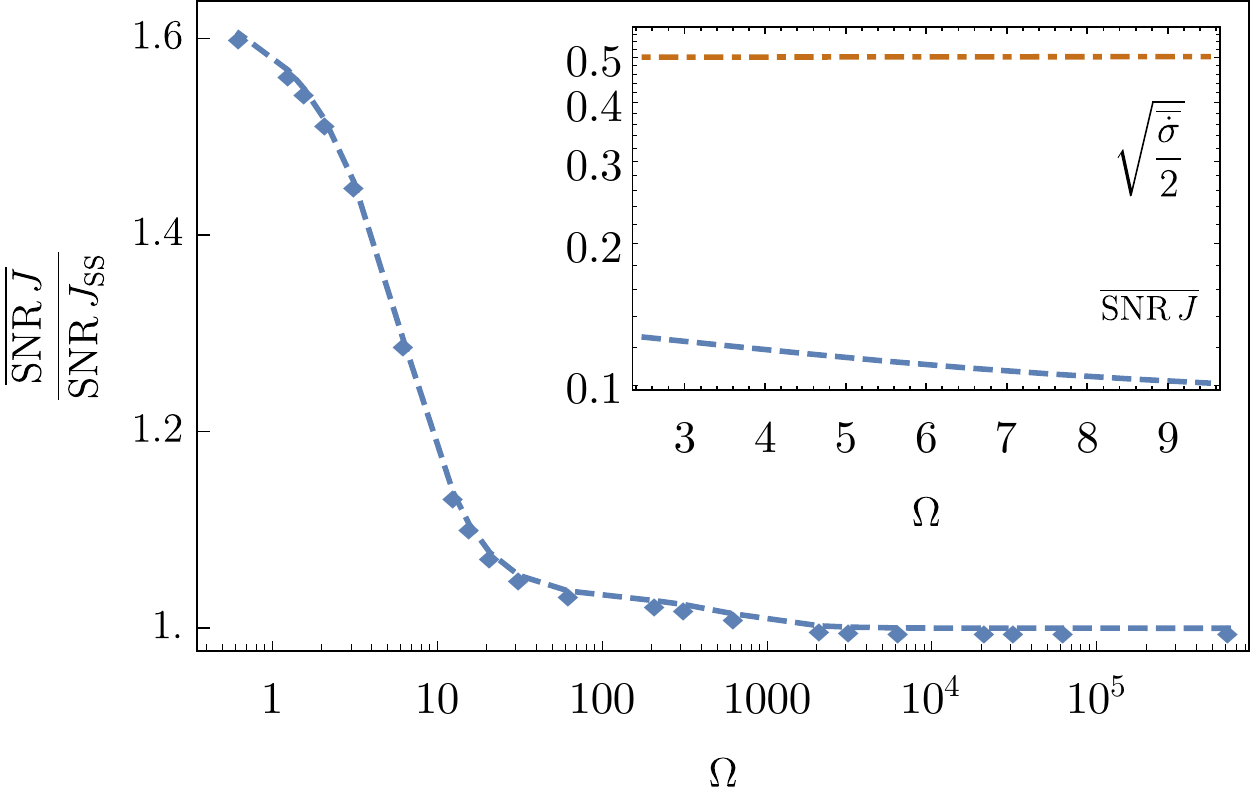}
  \caption{Fluctuations as a function of the driving frequency $\Omega$. Time-averaged, normalized signal-to-noise ratio $\overline{\text{SNR}\,J}/\text{SNR}\,J_{SS}$ as a function of  $\Omega$. \emph{Inset:} Time-averaged signal-to-noise ratio (dashed line) is loosely bounded above by the square root of half the dissipation per period $\overline{\dot \sigma}$  (dot-dashed line). Parameters for this plot are the same of Figure \ref{fig:scgf}, and $\gamma=1$.}
  \label{fig:ld_fluctuations}
  \label{fig:ld_snr}
\end{figure}
Finally, we are in the position to complete the analysis of the fluctuations as a function of the driving frequency. From the numerically computed scaled cumulant generating function \eqref{eq:periodicscgf}, we obtain the average variance and signal-to-noise ratio shown in Figure \ref{fig:ld_fluctuations}, normalized with respect to the corresponding steady-state results. The time-averaged variance is slightly peaked around $\Omega=0$,  and then decreases rapidly, before settling to the steady-state value at even higher frequencies.
An upper bound for the signal-to-noise ratio of a Markov process under symmetric driving is provided in terms of the entropy production averaged on a period \eqref{eq:epr}  \cite{fal19, proesmans2017discrete}.
As shown in the inset of  Figure~\ref{fig:ld_fluctuations}, the quantity 
\begin{align}
\sqrt{(\exp {\overline{\dot \sigma}}-1)/2} \simeq \sqrt{ \overline{\dot \sigma}/2} \label{eq:thermo_bound}
\end{align}
 bounds from above the time-averaged signal-to-noise ratio.

\section{Conclusion and perspectives}
\label{sec:discussion}

The central result of Section \ref{sec:reduction} is that a sinusoidal driving on the concentration of an externally regulated species, \emph{i.e.} a chemostat, leads to a non-zero contribution for the \emph{time-averaged} current that is maximum at small frequencies, under some conditions on the kinetic rates. This is analogous to stochastic resonance \cite{ben81, gammaitoni1998stochastic}, where a small time-periodic modulation of a system subject to noise leads to an effect that is dominant at small frequencies.We also discussed the effects on the fluctuations of the current using the quasi-steady-state approximation in the limit of vanishing driving frequency, and finally we framed the problem of computing the time-averaged cumulants in the language of large deviations for periodically driven systems, thus including the case of finite-frequency modulation.

Let us now reconsider the previous results in the light of their conceptual interest in the performance analysis of biochemical pathways in physiological conditions and, finally, of the possible applications in synthetic biology.

The biological purpose of substrate inhibition has been discussed in the review \cite{reed2010biological}. There, the authors considered various biologically significant examples where substrate inhibition can sizably enhance the stability of biological processes, for example limiting the effect of the environmental fluctuations on a target current. For the case of dopamine synthesis from tyrosine, they present numerical simulations that take into account the periodic variations in the substrate concentration. Reference \cite{falasco2019negative} reviewed the theoretical motivation behind this stabilization effect and showed, for the synthesis of serotonin out of tryptophane, that substrate inhibition can also maximize the signal-to-noise ratio of relevant reaction currents. Both these works used implicitly the quasi-steady-state approach presented in Section \eqref{sec:qss} when using the reaction velocity curve derived for the steady-state to discuss the stability to periodic variations of the chemostat concentrations on the scale of few hours. In our analysis, we found out that this approximation depends crucially on the choice of the kinetic constants and of the chemical affinities. However, when considering the choice of physiological parameters given in \cite{falasco2019negative} for serotonin synthesis out of tryptophan and for the dopamine synthesis from tyrosine, we see that the quasi-steady-state approach proves to be sufficient, and the frequency-dependend effect in Eq. \eqref{eq:current_correction} is negligible for every value of the frequency.

Finally, the principle explained in this work can be used to optimize the current response in engineered chemical systems. From Eq. \eqref{eq:optimization} is clear that the optimization problem to be solved is defined by the specific context in which the chemical pathway is placed, for example by the sign of the net current that we want to achieve. Since this is governed by the sign of the steady-state affinity \eqref{eq:affinity}, the natural optimization problem in this case is subject to a constraint on the sign of the latter, as it is illustrated in Figure \ref{fig:opt}. On the other hand, if one only cares about the magnitude of the current, regardless of its sign, the constraint on the sign of the affinity can relaxed. We note here that it is the presence of this constraint on $\mathcal{A}$ that, for the situation displayed in Figure \ref{fig:negativeDeltaJ}, causes the optimal value of $s$ to be located near $\mathcal{A}=0$. Generically, thermodynamics alone imposes loose constraints on the far from equilibrium behavior of even simple reaction networks \cite{baiesi2013update, england2015dissipative}. To further corroborate this idea, we have verified in the inset of Figure~\ref{fig:ld_fluctuations} that the system's dissipation per period \eqref{eq:epr}, despite providing the upper bound \eqref{eq:thermo_bound} on the signal-to-noise ratio \cite{fal19, proesmans2017discrete}, is a rather loose one, irrespective of the driving frequency. The goal of this optimization procedure, in the case of substrate-inhibition \eqref{eq:sub_inh}, can be to increase the average current, leading to an improved response to the external stimuli. This may be important for example to deplete the concentration of an unwanted species in a small time interval. On the other hand, the product-inhibition scheme \eqref{eq:prod_inhib} shows how it is possible to efficiently use the substrate S in a way that minimizes the total consumption (obtained as the time-averaged current $\mean{\overline{J}}$), while keeping the product concentration constant. Thus, coupling the kind of chemical pathways here analyzed to a recently proposed scheme of out-of-equilibrium synthesis \cite{penocchio2019thermodynamic} could lead to significant improvements.

We acknowledge funding from  the European Research Council project NanoThermo (ERC-2015-CoG Agreement No. 681456). DF thanks Emanuele Penocchio and Francesco Avanzini for fruitful discussions.
\newpage
\appendix

\widetext
\section{Current statistics and path integral approach to fluctuations}
\label{app:pathint}

Here we show that the statistics of the current $J$, measured on the reactions $\pm1$ or $\pm3$, is the same in the steady-state or in the periodic steady-state. In the steady-state case, the scaled cumulant generating function for the currents in the Markov process  can be obtained by taking the logarithm of the dominant eigenvalue of the tilted generator \eqref{eq:tiltedgen}. The latter is written considering explicitly the interesting current to be the one of reaction 3. However, one can ask if the results is changed by a different choice of the current, \emph{e.g.} current of reaction 1, that leads to the tilted generator
\begin{align}
\mathcal{T(\lambda)}=
  \begin{pmatrix}
    -(k_1s+k_{-3}p_0) & k_{-1}\e^{-\lambda}+k_3& 0 \\
    k_{-3}p_0 + k_1s \e^{\lambda}& -K & k_{-2}\\
    0 & k_2s & -k_{-2}\\ 
  \end{pmatrix}\,.
  \label{eq:tiltedgen2}
\end{align}
We show that the characteristic equation for the eigenvalues of the generator is invariant under this operation. In fact, the equation for the eigenvalues $\mu$, written using the generator \eqref{eq:tiltedgen}, reads
\begin{align}
  & (\mu +  k_1s+k_{-3}p_0) (\mu + K) + (k_{-1}+k_3\e^{\lambda})(k_{-3}p_0 \e^{-\lambda} + k_1s) \nonumber\\
  & =k_2s (\mu +  k_1s+k_{-3}p_0) \,. \label{eq:characteristic}
\end{align}
The counting field only appears in the second term of the first line, and it is evident that collecting the exponentials in the two factors leads to the same equation that would derive from \eqref{eq:tiltedgen2}.

To generalize this result to the case of a periodic steady-state, we start from the generating function $g(\lambda) $ of the current $ J_\alpha$ along reaction $\alpha$, written as the path integral \cite{falasco2019negative, laz19}
\begin{align}
g_\alpha(\lambda) =  \int \mathcal{D}n \mathcal{D}\Pi \, e^{\int_0^T dt \left[ -\Pi \cdot \dot n +H_\alpha(n,\Pi,\lambda) \right]}\label{cgf}
\end{align}
with effective `Hamiltonian' 
\begin{align}
H_\alpha(n,\Pi,\lambda):= \sum_{\rho} W_{\rho}(n)(e^{\Pi \cdot \nu +  o_\rho \lambda }-1).
\end{align}
Here $n(t)$ is the vector of the instantaneous species number (with $\Pi(t)$ its conjugated variable), $W_{\rho}$ is the rate of reaction $\rho$, and $\nu_{\rho i}$ the stoichiometric matrix giving the variation of a species $i$ due to a reaction $\rho$. Here $o_\alpha=-o_{-\alpha}=1$ and $o_\rho=0, \, \forall \rho \neq \alpha$.

In the large time $T$ limit, the functional integral in \eqref{cgf} is dominated by the trajectories $n^*_i$ and $\Pi^*_i$ that maximize the exponential, that are the solutions of the equations of motion
\begin{align}
 \begin{aligned}\label{eom}
\dot n_i &=   \sum_{\rho}  \nu_{\rho i}W_{\rho}(n) e^{\Pi \cdot \nu +  o_\rho \lambda } ,\\
\dot \Pi_i &=-  \sum_{\rho}\partial_{n_i} W_{\rho}(n)(e^{\Pi \cdot \nu +  o_\rho \lambda }-1) .
 \end{aligned}
\end{align}

Within this formalism, the equality of the long-time statistics of reaction 3 and 1 can be obtained as follows. Consider the change of variables $\Pi_{\text{E}}\to\Pi_{\text{E}}+ \lambda$ and $\Pi_{\text{ES}}\to\Pi_{\text{ES}}- \lambda$  in \eqref{cgf}. It transforms the Hamiltonian  $H_1(n,\Pi,\lambda)$ into $H_3(n,\Pi,\lambda)$ and produces the extra term $e^{ - \lambda \int_0^T dt (\dot n_\text{E}-\dot n_\text{ES})}$. The latter, when evaluated on the dominant trajectories \eqref{eom}, is either unity in a stationary state, or sub-extensive in $T$ in a periodic state (due to the periodicity of $n(t)$). In other words, we regained that (for large $T$) $ g_1(\lambda)= g_3(\lambda)$ in a stationary state, and that
\begin{align}\label{g1g3}
\ln g_1(\lambda) = \ln g_3(\lambda) + h.o.t
\end{align}
in a periodic state, where the higher order terms drop out when rescaling \eqref{g1g3} to obtain \eqref{eq:periodicscgf}.

The path-integral formalism is convenient to describe the relation between the computation of the complete scaled cumulant generating function, the quasi-steady-state approximation and the geometric contribution derived in \cite{sin07}.  If reaction rates do not depend on time (in our case, when the product concentration is kept constant), the dominant solution to Eqs. \eqref{eom} will be given by the solution of the \emph{stationary} Hamilton equations obtained setting $\dot{n}=\dot{\Pi}=0$ and solving the system while imposing the conservation laws 
\begin{align}
  \sum_i n_i=\frac{M}{V}\,,\\
  \sum_i \Pi_i =0 \,.
\end{align}
Under this condition, the kinetic contribution to the action functional  drops, since it contains the time derivative $\dot n $.

When allowing a slow variation of the vector of the reaction rates $k$, the stationary Hamilton equation will continue to give the dominant contribution for every fixed value of $k$, but the action functional should be computed taking into account the parametric dependence on time of the vector $k=k(t)$.
In that case the contributions of the Hamiltonian part and of the kinetic part of the action can rewritten respectively as
\begin{align}
  \int_0^T H_\alpha(n^*(k(t)), \Pi^*(k(t)),\lambda) dt \label{qss-contr}
\end{align}
and
\begin{align}
\int_0^T \Pi^*(k(t)) \cdot  \dot{n}^*(k(t)) d t = \oint \Pi^*(k) \cdot dn^*(k)\,.\label{geom-contr}
\end{align}
The QSS approach considers the term in Eq. \eqref{qss-contr} as the only contribution to the scaled cumulant generating function, neglecting the geometric contribution given by Eq. \eqref{geom-contr}. Note that the latter term gives a contribution to the cumulants of the time-averaged current that is proportional to the driving frequency.
\bibliography{./bibliography} 
\end{document}